\documentclass{PoS}

\title{Global reanalysis of nPDFs}

\ShortTitle{Global reanalysis of nPDFs}

\author{\speaker{K. J. Eskola}\thanks{This work was financially supported the Academy of Finland, %%@
Projects 73101, 80385, 206024 and 115262.}, 
		 V. J. Kolhinen, H. Paukkunen\\

        Department of Physics, University of Jyv\"askyl\"a and 
		Helsinki Institute of Physics, Finland \\

        E-mail: \email{kari.eskola,vesa.kolhinen,hannu.paukkunen@phys.jyu.fi}}		

\author{C.A. Salgado\thanks{Financially supported by the 6th Framework Programme 
		 of the European Community under the Marie Curie contract MEIF-CT-2005-024624.}\\

        Dipartimento di Fisica, Universit\`a  di Roma ``La Sapienza'' and INFN, 
Rome, Italy \\

        E-mail: \email{carlos.salgado@cern.ch}}

\abstract{In this talk, we present the results from our recent global reanalysis of nuclear parton %%@
distribution functions (nPDFs), where the DGLAP-evolving nPDFs are constrained by nuclear hard %%@
process data from deep inelastic $l+A$ scattering (DIS) and the Drell-Yan (DY) process in $p+A$ %%@
collisions, and by sum rules. The main improvements over our earlier work {\em EKS98} are the %%@
automated $\chi^2$ minimization, better controllable fit functions and possibility for error %%@
estimates. The obtained 16-parameter fit to $N=514$ datapoints is good, $\chi^2/{\rm d.o.f}=0.82$. %%@
Fit quality comparison and the error estimates obtained show that the old {\em EKS98} parametrization %%@
is fully consistent with the present automated reanalysis. Comparison with other global nPDF analyses %%@
is presented as well. Within the DGLAP framework we also discuss the possibility of incorporating a %%@
clearly stronger gluon shadowing, which is suggested by the RHIC BRAHMS data from d+Au collisions.
}

\FullConference{High-$p_T$ physics at LHC\\

		 March 23-27, 2007 \\

		 University of Jyv\"askyl\"a, Jyv\"askyl\"a, Finland}

\begin{document}
\section{Introduction}
Inclusive cross sections of hard processes in high energy hadronic and nuclear collisions are %%@
computable using the collinear factorization theorem of QCD, 
$
\sigma_{AB\rightarrow h+X} = \sum_{i,j} f_{i}^A(x_1,Q^2) \otimes f_{j}^B(x_2,Q^2) \otimes
\sigma_{ij\rightarrow h+X}.
$
In these computations the short-distance pieces, the squared perturbative QCD (pQCD) 
matrix elements, are contained in $\sigma_{ij\rightarrow h+X}$. The long-distance nonperturbative %%@
input is included in the universal process-independent parton distribution functions (PDFs), %%@
$f_{i}^A(x,Q^2)$, which depend on the momentum fraction $x$ carried by the colliding parton $i$,
the factorization scale $Q$, and the type $A$ of the colliding hadron or nucleus.
Once the PDFs are known at an initial scale $Q_0\gg \Lambda_{\rm QCD}$, the DGLAP equations %%@
\cite{DGLAP} predict their behaviour at other (perturbative) scales. 

In the global analysis of PDFs, the goal is to determine the DGLAP-evolving PDFs in a %%@
model-independent way on the basis of constraints offered by the sum rules for momentum, baryon %%@
number and charge conservation, and in particular by multitude of hard QCD-process data in hadronic %%@
and nuclear collisions. The global analysis becomes, however, cumbersome already on hadronic level, %%@
since the data from which the PDFs are to be determined do not lie along constant scales $Q^2$ in the %%@
$(x,Q^2)$ plane. Furthermore, the kinematical ranges of the data from different measurements often do %%@
not overlap and the precision of the data may vary.
For these reasons, the  global analysis of the PDFs usually proceeds with the following 
steps: 
\emph{(i)} Choose a suitably flexible functional form for the PDFs, expressed in terms of several %%@
enough but not too many parameters at an initial scale $Q_0\sim 1$~GeV. Use sum rules to reduce the %%@
number of parameters.
\emph{(ii)} Evolve the PDFs to higher scales according to the DGLAP equations.
\emph{(iii)} Compare with the hard process data available, compute overall $\chi^2$ to quantify the %%@
quality of the obtained fit. 
\emph{(iv)} Iterate the initial parameter values until a best (local) minimum of $\chi^2$ in the 
multi-dimensional parameter space, a best fit, is found.
 
The global analyses carried out by the MRS group  \cite{MRS03} and by the CTEQ collaboration  
\cite{CTEQ6} have been quite successful in pinning down the PDFs of the free proton at the 
next-to-leading order (NLO) level of pQCD, and the analysis there is now moving already 
to the NNLO level. For the \emph{nuclear} PDFs (nPDFs), three groups 
have so far presented results from a global analysis: 
\begin{itemize}
\item \emph{EKS98} \cite{Eskola:1998iy,Eskola:1998df} was the first global analysis performed for the %%@
nPDFs. This leading-order (LO) analysis demonstrated that the measured cross sections for deep %%@
inelastic lepton-nucleus scattering (DIS) and for the Drell-Yan (DY) dilepton production in %%@
proton-nucleus collisions, and in particular the $\log Q^2$-slopes of $F_2^{\rm Sn}/F_2^{\rm C}$ can %%@
all be reproduced and the momentum and baryon number sum rules required simultaneously within the %%@
DGLAP framework. The original data fitting in \emph{EKS98} was, however, done by eye only. 
%As described below, we have now in \cite{EKPS} improved upon this feature and obtained also %%@
%statistical uncertainty limits for the nuclear effects.

\item \emph{HKM} \cite{Hirai:2001np} and \emph{HKN} \cite{Hirai:2004wq} were the first nPDF global %%@
analyses with $\chi^2$ minimization automated and also uncertainties estimated. The nuclear DY data %%@
were not included in \emph{HKM} but were added in \emph{HKN}. These analyses were still at the LO %%@
level.

\item \emph{nDS} \cite{deFlorian:2003qf} was the first NLO global analysis for the nPDFs.  
\end{itemize}

The main goals of the global reanalysis of nPDFs which we have recently performed in 
\cite{EKPS} and discuss in this talk, can be summarized as follows:
As the main improvement over the \emph{EKS98}, we now automate the $\chi^2$ minimization.
We check whether the already good fits obtained in \emph{EKS98} could yet be improved. 
We now also report uncertainty bands for the EKS98-type nuclear effects of the PDFs. 
We also want to check whether the DIS and DY data could allow stronger gluon shadowing than 
obtained in \emph{EKS98}, \emph{HKN} and \emph{nDS}. The motivation for this are the 
BRAHMS data \cite{Arsene:2004ux} for inclusive hadron production in d+Au collisions at RHIC 
which show a systematic suppression relative to p+p at forward rapidities.

\section{The framework}

We define the nPDFs, $f_{i}^A(x,Q^2)$, as the PDFs of bound \emph{protons}. By nuclear modifications, %%@
$R_{i}^A(x,Q^2)$, we refer to modifications relative to the free proton PDFs, 
\begin{equation}
f_{i}^A(x,Q^2) = R_{i}^A(x,Q^2) f_{i}^{\rm CTEQ6L1}(x,Q^2),
\end{equation}  
which in turn are here supposed to be fully known and which are taken from the CTEQ6L1 set %%@
\cite{CTEQ61}.
Above, $i$ is the parton type and $A$ is the mass number of the nucleus. The PDFs of the bound %%@
neutrons are obtained through isospin symmetry ($u^{n/A}=d^{p/A}$ etc.), which is exact for isoscalar %%@
nuclei and assumed to hold also for the non-isoscalar nuclei. The initial scale is the lowest scale %%@
of the CTEQ6L1 disctributions, $Q_0=1.3$~GeV. The small nuclear effects for deuterium are neglected.

As in \emph{EKS98}, we consider only three different modifications at $Q_0$: 
$R_V^A$ for valence quarks, $R_S^A$ for all sea quarks, 
and $R_G^A$ for gluons. Further details cannot, unfortunately,  
be specified simply due to the lack of data. Each of these ratios consists of three pieces,
which are matched together at the antishadowing maximum at $x=x_a^A$ and at the EMC minimum at %%@
$x=x_e^A$ (cf. Fig.~\ref{Fig:Initial}):
\begin{eqnarray}
  &&R_1^A(x) = c_0^A+(c_1^A+c_2^A x)[\exp(-x/x_s^A) 
           - \exp(-x_a^A/x_s^A)],   \qquad x\le x_a^A \label{R1}\\
  &&R_2^A(x) = a_0^A + a_1^A x + a_2^A x^2 + a_3^A x^3, 
             \qquad x_a^A \le x \le x_e^A\\ \label{R2} 
  &&R_3^A(x) = \frac{b_0^A-b_1^A x}{(1-x)^{\beta_A}},      
             \qquad x_e^A \le x \label{R3}.
\end{eqnarray}

As explained in \cite{EKPS}, we convert the parameters into the following
more transparent set of seven parameters, 

\begin{tabular}{ll}
  $y_0^A$          & $R_1^A$ at $x\rightarrow 0$, defining where shadowing levels off,\\
  $x_s^A$          & a slope factor in the exponential,\\
  $x_a^A$, $y_a^A$ & position and height of the antishadowing maximum,\\
  $x_e^A$, $y_e^A$ & position and height of the EMC mimimum,\\
  $\beta_A$		   & slope of the divergence of $R_3^A$ caused by Fermi motion at $x\rightarrow 1$.
\end{tabular}
\vspace{0.2cm}

\noindent The $A$-dependence of nPDFs is contained in the $A$-dependence of each parameter,
taken to be of the following simple 2-parameter form,
\begin{equation}
  z_i^A = z_i^{A_{\rm ref}} (\frac{A}{A_{\rm ref}})^{\,p_{z_i}},
  \label{eq:Adependence}
\end{equation}
where $z_i = x_s, x_a, y_a\ldots$, and where Carbon ($A_{\rm ref}=12$) is chosen  as the reference %%@
nucleus.

To reduce the number of parameters from $3\times 14=42$ down to our final set of 16 free parameters %%@
to be determined by $\chi^2$ minimization with the MINUIT routine \cite{MINUIT}, we impose
baryon number and momentum conservation, and fix the initial large-$x$ gluon and sea quark %%@
modifications (which in practice remain unconstrained) to $R_V^A$. Lots of manual labour was still %%@
required for finding converging fits, starting values and ranges for the 16 free fit parameters.

\section{Results}
The data sets against which the best fit was found, are 
the DIS data from 
NMC \cite{Arneodo:1995cs,Amaudruz:1995tq,Arneodo:1996rv,Arneodo:1996ru}
FNAL E665 \cite{Adams:1995is} and 
SLAC E-139 \cite{Gomez:1993ri}, 
and the DY data from 
FNAL E772 \cite{Alde:1990im} and 
FNAL E866 \cite{Vasilev:1999fa}. 
For the available $A$-systematics and other details, consult Table 1 in %%@
\cite{EKPS}\footnote{Correction: NMC 96 data for Sn/C, used in our reanalysis, should appear in Table %%@
1 of \cite{EKPS} as well.}.

The obtained parameters corresponding to the best fit found are shown in Table
\ref{Table:Params}. The goodness of the fit was $\chi^2 = 410.15$ for $N=514$
data points and 16 free parameters, which corresponds to $\chi^2/N=0.80$ and 
$\chi^2/\mathrm{d.o.f.} = 0.82$. 

\begin{table}[tbh]
\begin{center}
\begin{tabular}{l|c|lll}
   & Param.   	&  Valence      	&  Sea           	&  Gluon \\
\hline
\hline
 1 &  $y_0$    	& {baryon sum}	&  0.88909       	&  {momentum sum}	\\
 2 &  $p_{y_0}$ & {baryon sum}	&   -8.03454E-02  	& {momentum sum}  	\\ 
 3 &  $x_s$    	&  0.025 ($l$) 		&  0.100 ($u$)	&  0.100 ($u$)   \\
 4 &  $p_{x_s}$ &  0, {fixed}   &  0, {fixed}   &  0, {fixed}   \\
 5 &  $x_a$    	&  0.12190      	&  0.14011       	&  {as valence} \\
 6 &  $p_{x_a}$ &  0, {fixed}   &  0, {fixed}   &  0, {fixed}   \\
 7 &  $x_e$    	&  0.68716      	&  {as valence} &  {as valence} \\
 8 &  $p_{x_e}$ &  0, {fixed}   &  0, {fixed}   &  0, {fixed}   \\
 9 &  $y_a$    	&  1.03887      	&  0.97970       	&  1.071 ($l$)    \\
 10&  $p_{y_a}$ &  1.28120E-2   	& -1.28486E-2   	&  3.150E-2 ($u$)  		\\
 11&  $y_e$    	&  0.91050      	&  {as valence} &  {as valence} \\
 12&  $p_{y_e}$ & -2.82553E-2   	&  {as valence} &  {as valence} \\
 13&  $\beta$  	&  0.3          	& {as valence}  &  {as valence} \\
 14& $p_{\beta}$&  0, {fixed}   & {as valence}  &  {as valence}	\\
\hline
\end{tabular}                                                  \\
($u$) upper limit; ($l$) lower limit; 

\caption[]{\small The obtained final results for the free and fixed parameters defining the initial %%@
modifications $R_V^A$, $R_S^A$ and $R_G^A$ at $Q_0^2=1.69$~GeV$^2$. The powers $p_i$ define the %%@
$A$-dependence in the form of Eq.~(\ref{eq:Adependence}), the other parameters are for the reference %%@
nucleus $A=12$. Parameters which drifted to their upper (u) or lower (l) limits are indicated, see %%@
\cite{EKPS} for details. 
}
\label{Table:Params}
\end{center}
\vspace{-0.5cm}
\end{table}

\begin{figure}[hbt]
\centering\includegraphics[width=12cm]{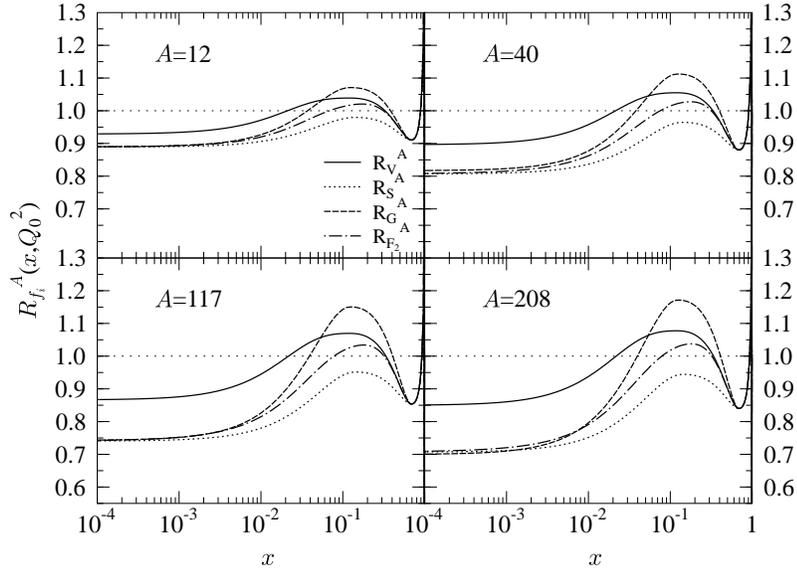}
\vspace{-0.5cm}
\caption[]{\small Initial nuclear modifications $R_V^A$ (solid lines),
  $R_S^A$ (dotted lines), $R_G^A$ (dashed lines) and
  $R_{F_2}^A = \frac{1}{A}{F_2}^A/\frac{1}{2}{F_2}^D$ 
  (dotted-dashed lines) for $A=12$, 40, 117 and 208
  as a function of $x$ at $Q_0^2=1.69$ GeV$^2$.}
\label{Fig:Initial}
\end{figure}

\begin{figure}[htb]
\vspace{-1cm}
\centerline{\includegraphics[width=8cm]{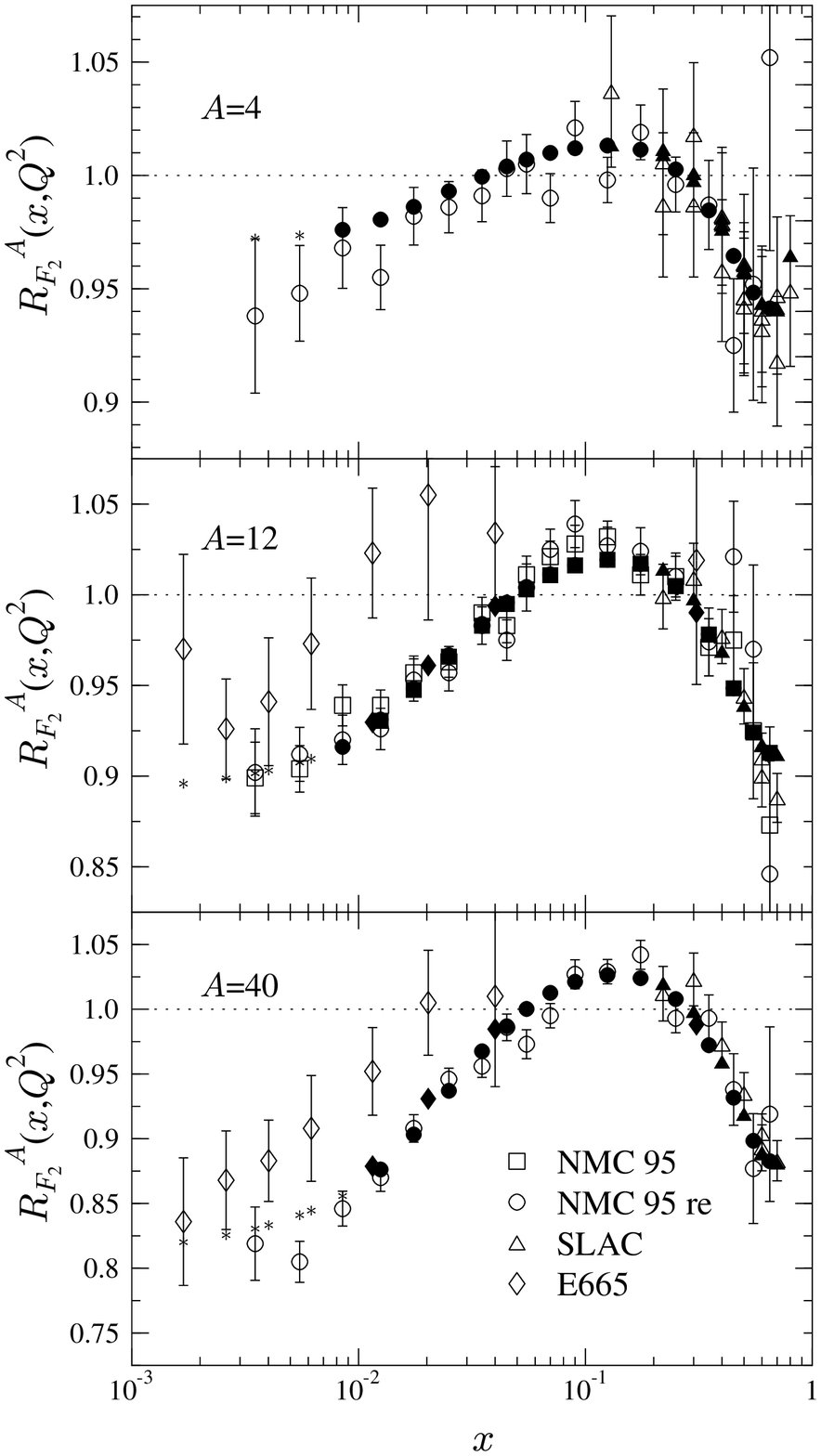} %%@
\includegraphics[width=9cm]{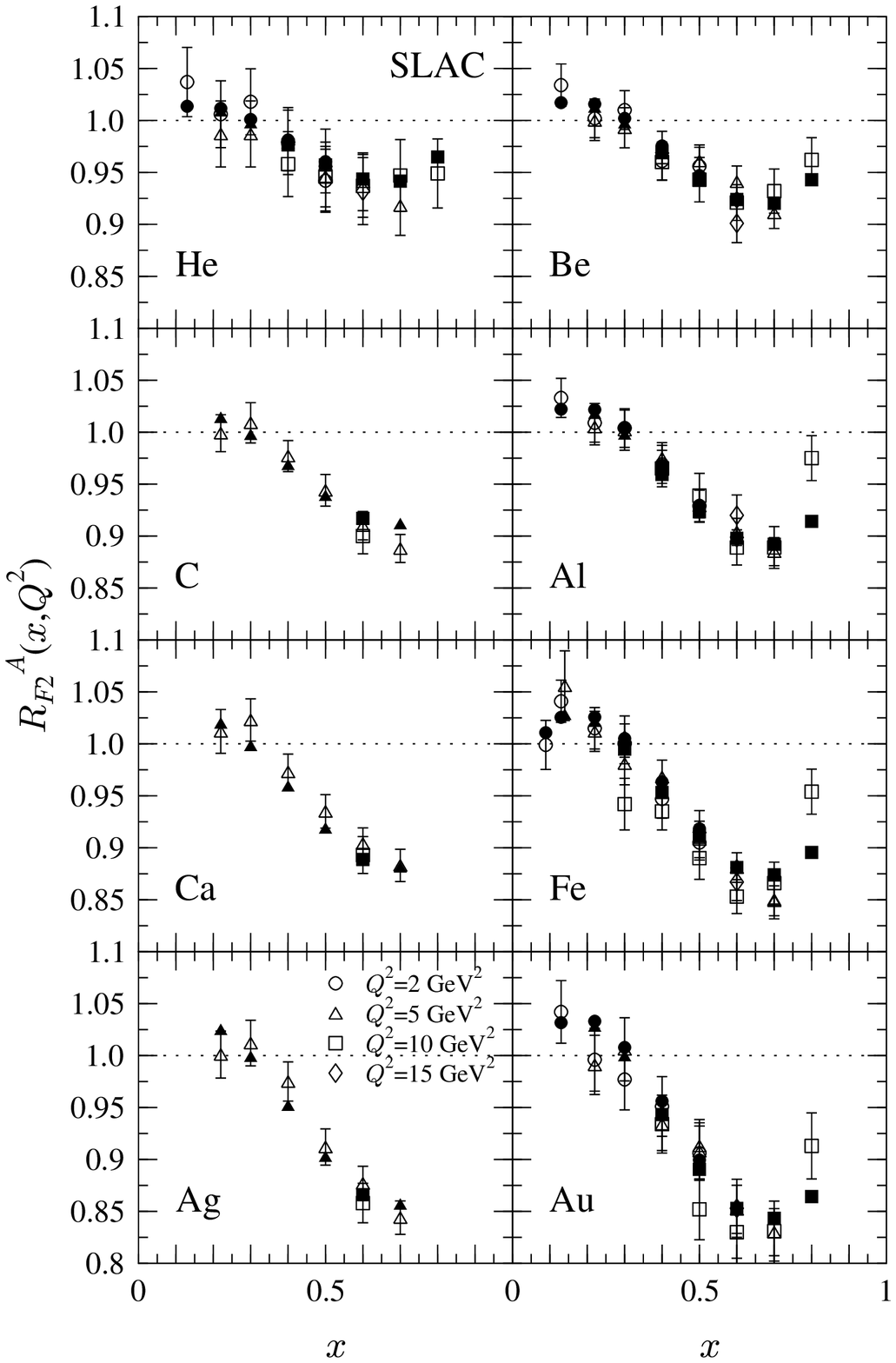}}
\vspace{-0.5cm}
\caption[]{\small \textbf{Left:} The calculated %%@
$R_{F_2}^A(x,Q^2)=\frac{1}{A}{F_2}^A/\frac{1}{2}{F_2}^D$ (filled symbols)
  against the data from SLAC  E-139 (triangles) \cite{Gomez:1993ri}, E665 (diamonds)
  \cite{Adams:1995is} and NMC (squares and circles)
  \cite{Arneodo:1995cs,Amaudruz:1995tq}.  The asterisks denote our results
  calculated at $Q_0^2$ when $Q^2$ of the data is below $Q_0^2$.
  \textbf{Right:} Comparison with the SLAC  E-139 data  \cite{Gomez:1993ri} at 
  different fixed scales.}
\label{Fig:RF2A1}
\end{figure}

\begin{figure}[htb]
\vspace{-.0cm}
\centerline{\includegraphics[width=8cm]{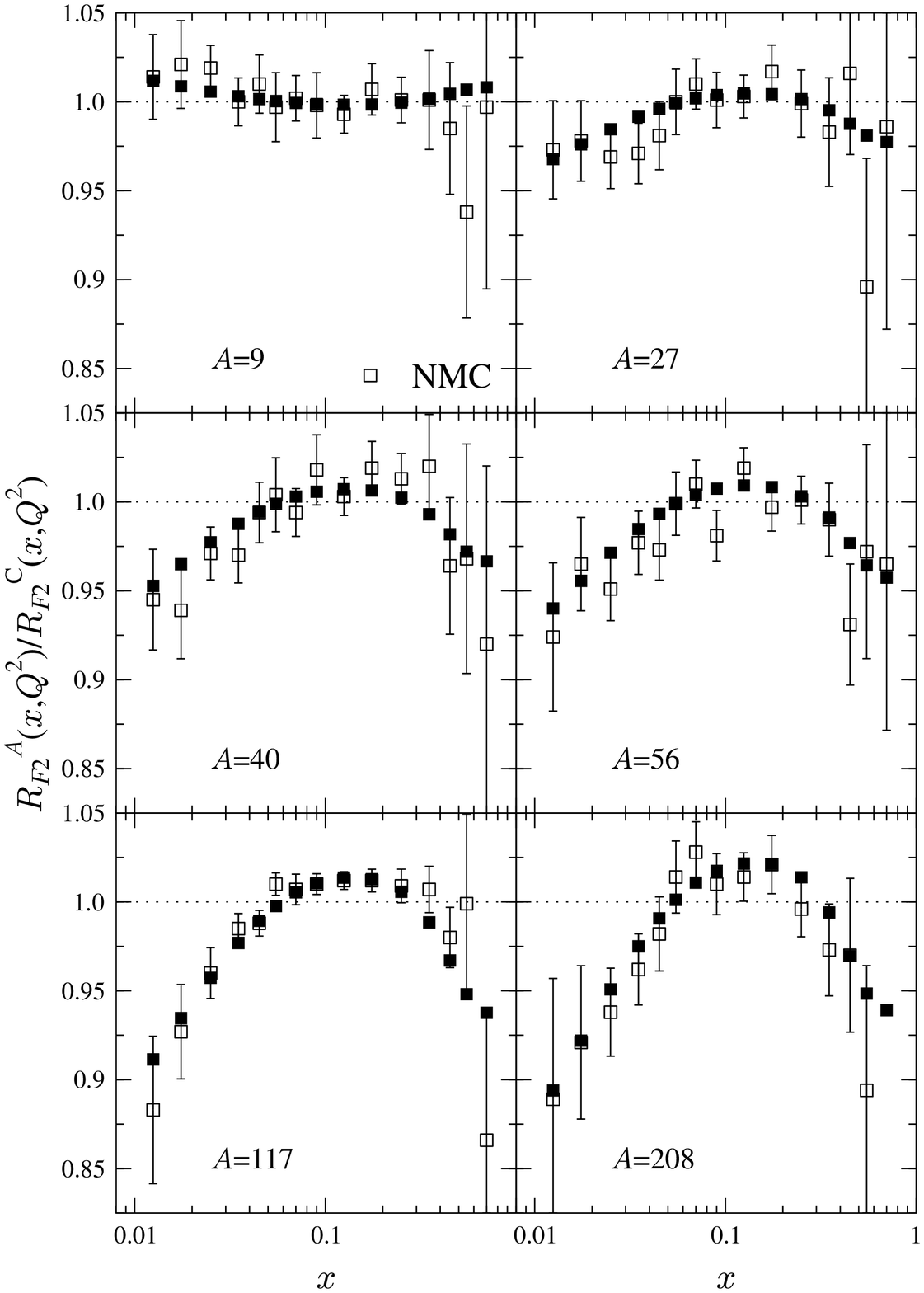} \includegraphics[width=8cm]{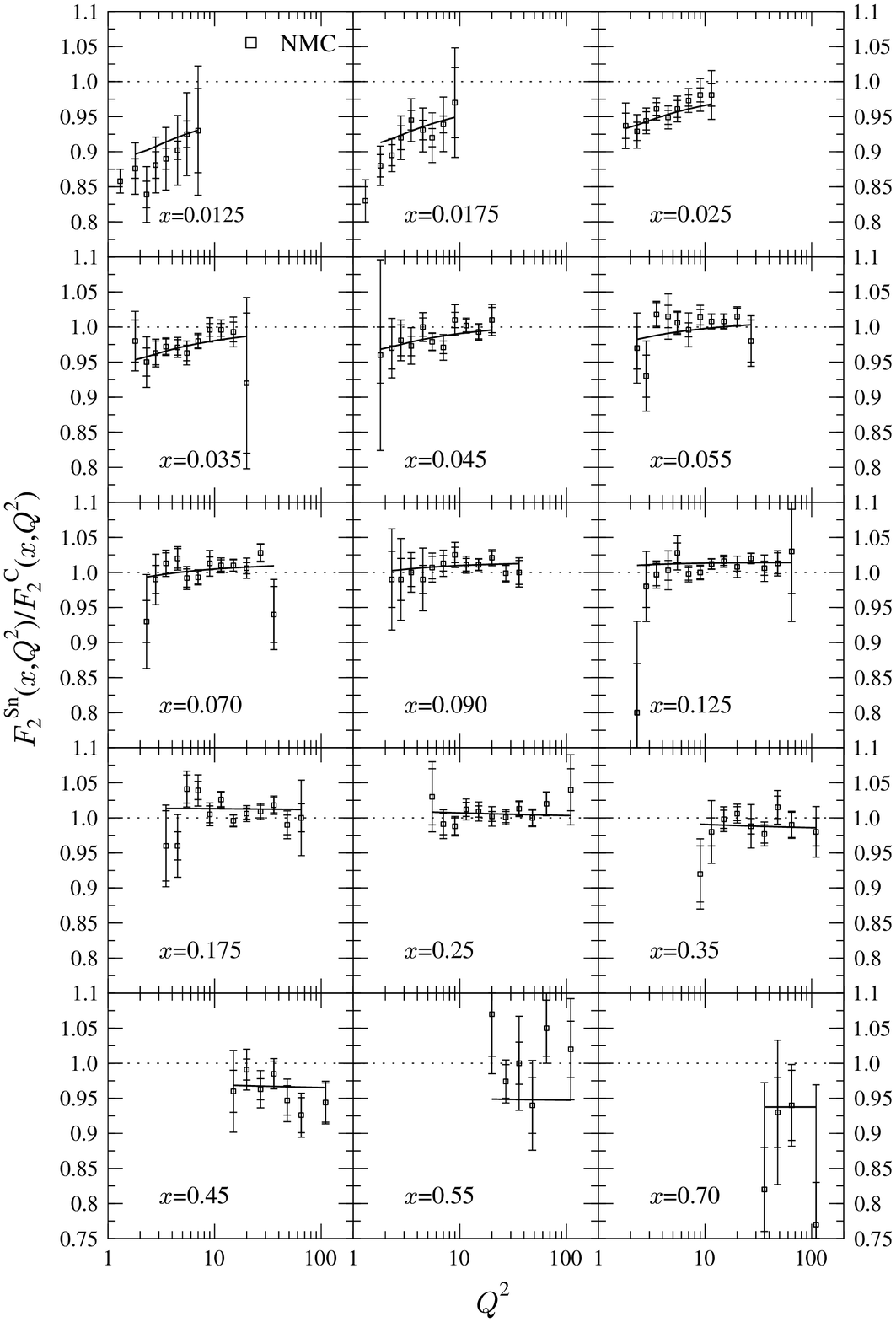}}
\vspace{-0.3cm}
\caption[]{\small 
\textbf{Left:} The computed ratios $\frac{1}{A}F_2^A/\frac{1}{12}F_2^{\mathrm C}$ 
(filled squares) and the NMC data \cite{Arneodo:1996rv} (open squares).
\textbf{Right:} The calculated scale evolution (solid lines) of the ratio
  $F_2^{\mathrm{Sn}}/F_2^{\mathrm{C}}$ against the NMC data
  \cite{Arneodo:1996ru} for  various fixed values of $x$. The inner error bars are the statistical %%@
ones, the outer ones represent the statistical and systematic errors added in quadrature.
}
\label{Fig:RF2AC}
\vspace{-0.2cm}
\end{figure}

\begin{figure}[htb]
\vspace{1cm}
\includegraphics[width=8cm]{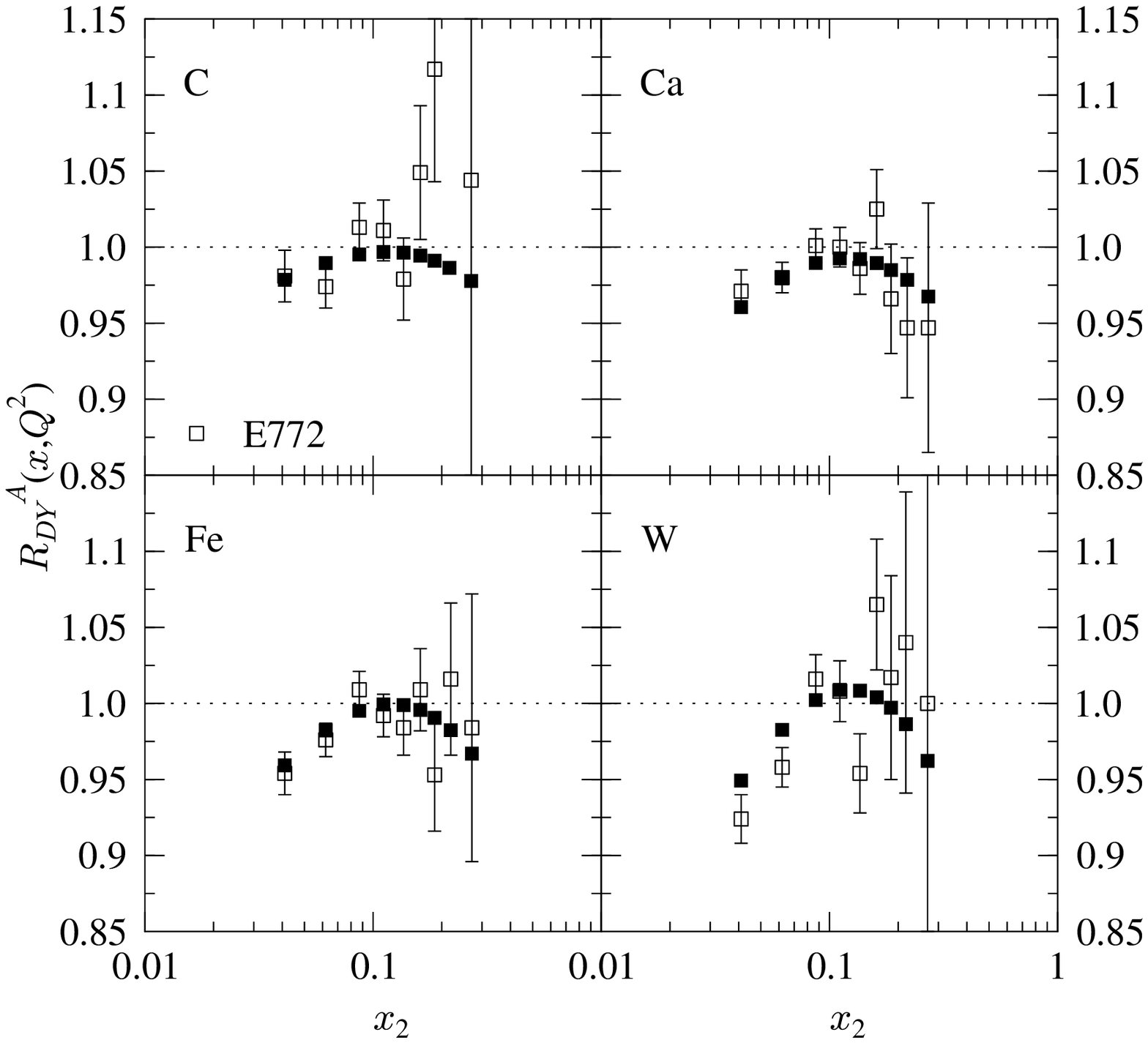} \hfill
 
\vspace{-9cm}\hspace{7.5cm}\includegraphics[width=9cm]{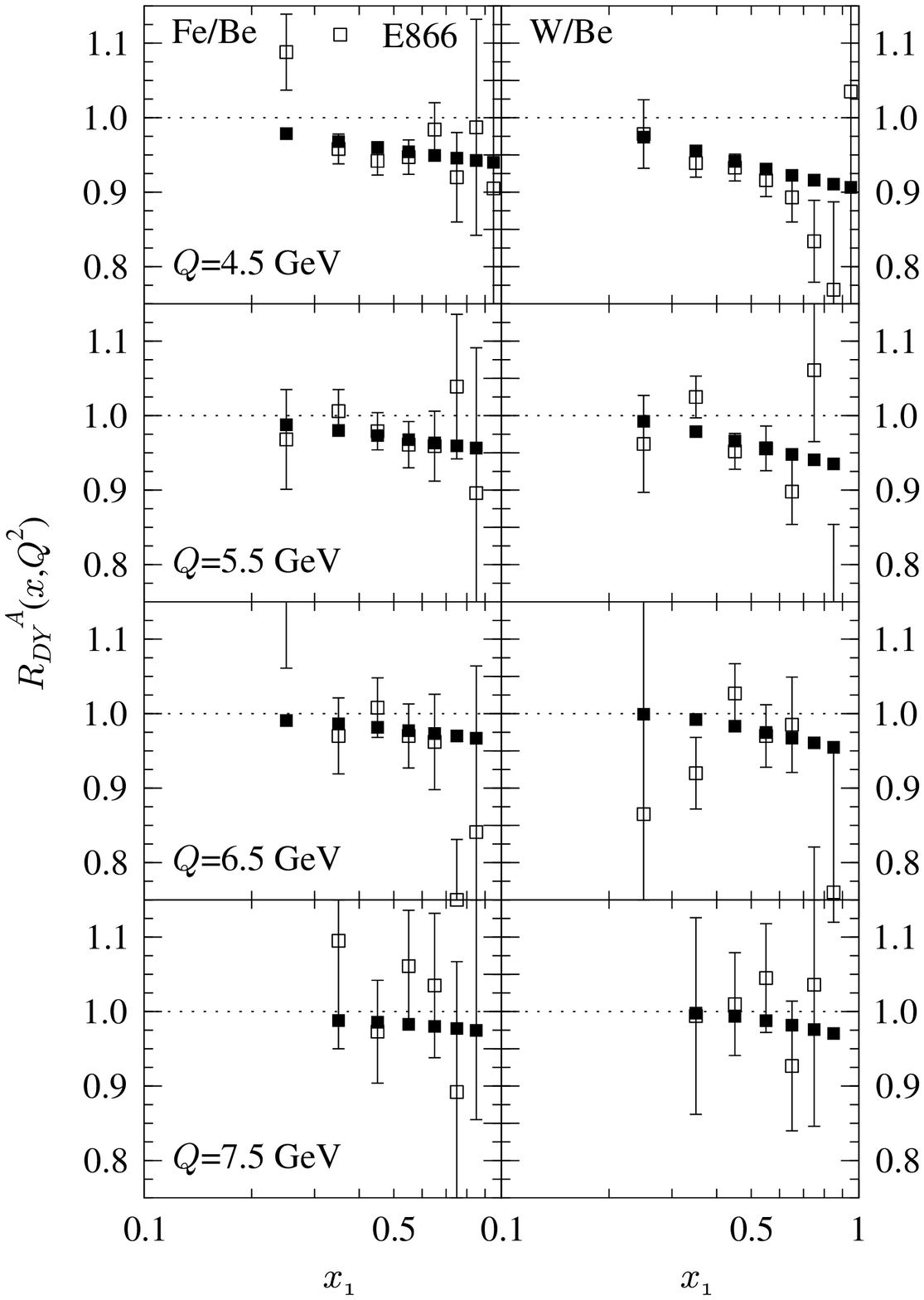}

\vspace{-0.7cm}
\caption[]{\small \textbf{Left:} The computed LO DY ratio (filled squares)
  $(d\sigma^{{\mathrm p}A}/dQ^2dx_2)/(d\sigma^{\mathrm{pD}}/dQ^2dx_2)$ against 
  the E772 data \cite{Alde:1990im} (open squares).
  \textbf{Right:} The computed LO DY ratio (filled squares)
  $(d\sigma^{{\mathrm p}A}/dQ^2dx_1)/(d\sigma^{\mathrm{pD}}/dQ^2dx_1)$ compared with the 
  E866 data  \cite{Vasilev:1999fa} (open squares) as a function of $x_1$ at four different 
  invariant-mass ($Q^2$)  bins.
}
\label{Fig:E772_E866}
\vspace{-0.2cm}
\end{figure}

The obtained initial nuclear modifications at $Q_0^2=1.69$ GeV$^2$ are shown in %%@
Fig.~\ref{Fig:Initial} for selected nuclei. Figs.~\ref{Fig:RF2A1}-\ref{Fig:E772_E866} show the %%@
obtained good agreement with the DIS and DY data. The computed results are shown with filled symbols, %%@
the data with open ones. For further details, consult the figure captions.

%The $A$-systematics obtained from DIS is shown in Fig. \ref{Fig:RF2AC} where the computed ratio 
%$\frac{1}{A}F_2^A/\frac{1}{12}F_2^{\mathrm C} = R_{F_2}^A/R_{F_2}^{\mathrm C}$ 
%is plotted against the  NMC data \cite{Arneodo:1996rv} for various nuclei. 

\begin{figure}[htb]
\vspace{-1cm}
\centering\includegraphics[width=12cm]{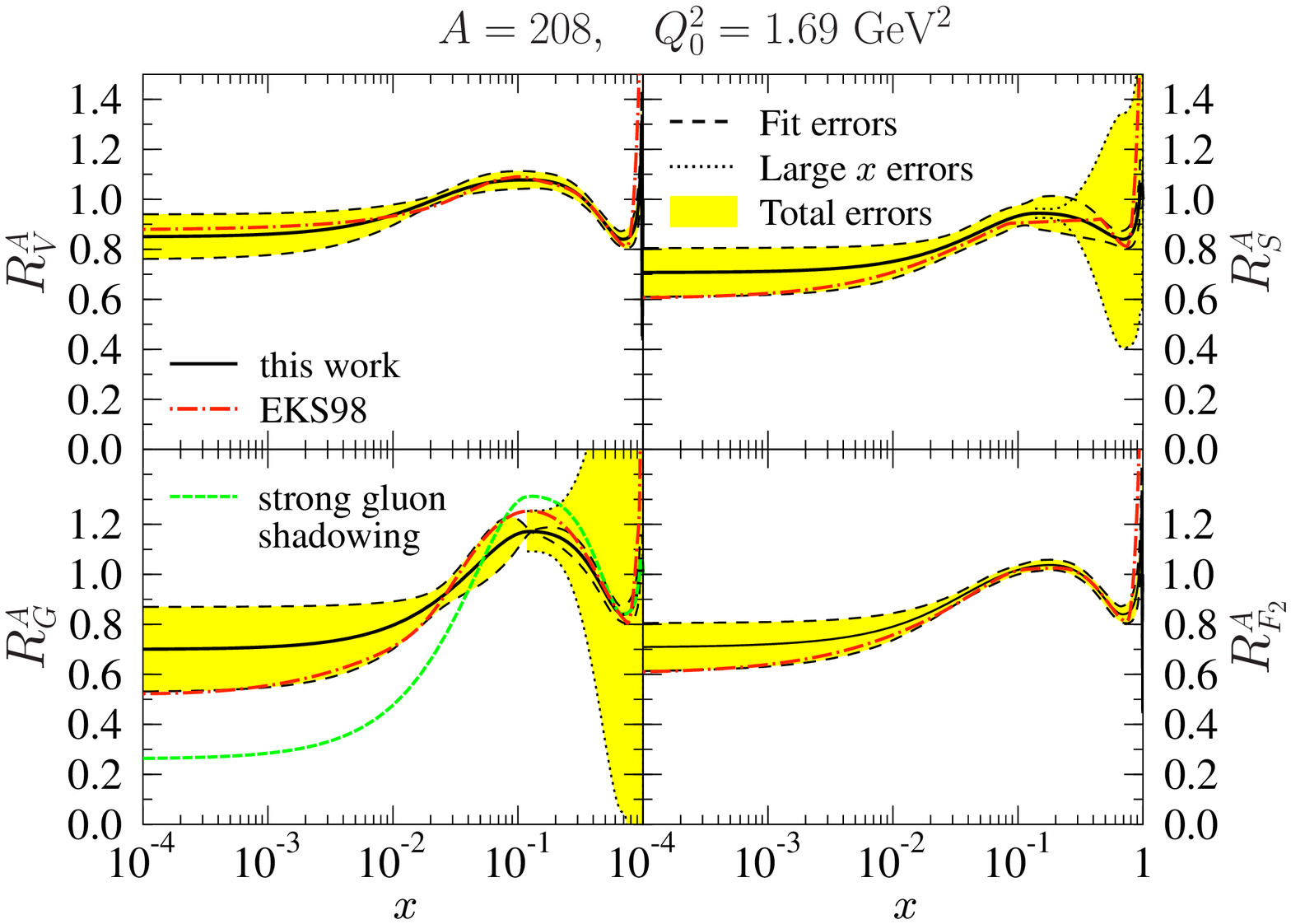}
\caption[]{\small  Error estimates for $R_i^A(x,Q_0^2)$ for Lead.
  Fit errors are shown by the dashed lines.  For large-$x$ sea quark and gluon modifications 
  the errors shown by the dotted lines were calculated separately, and 
  the yellow bands are the total error estimates obtained (see the text).
  The {\em EKS98} results, correspondingly evolved downwards from $Q_{0,EKS}^2=2.25$~GeV$^2$, 
  are shown by the dot-dashed red lines.  An example of a stronger gluon shadowing case is shown by 
  the dense-dashed green line.
  }
\label{Fig:Errors}
\vspace{-0.5cm}
\end{figure}

We obtain uncertainty estimates for the initial nuclear modifications using the Hessian error matrix %%@
output provided by MINUIT (for details and refs., see again \cite{EKPS}). These bands are denoted as %%@
"Fit errors" in Fig.~\ref{Fig:Errors}. To obtain physically more relevant large-$x$ errors for %%@
$R_G^A$ and $R_S^A$, which in the $\chi^2$ analysis were fixed to $R_V^A$ at large-$x$, we keep their %%@
small-$x$ parameters fixed and release the large-$x$ parameters for each $R_i^A$ at the time. This %%@
results in the "Large-$x$ errors" shown in Fig.~\ref{Fig:Errors}. The estimated total errors are then %%@
the yellow bands. 

\section{Conclusions from global reanalysis}

The total error bands in Fig.~\ref{Fig:Errors} demonstrate where and to what extent the available DIS %%@
and DY data constrain the nuclear modifications: the valence quarks average modifications $R_V^A$ are %%@
rather well, and independently of the functional form chosen, under control over the whole $x$-range, 
and so are the sea quarks at $x\sim 0.01-0.1$. At larger $x$, sea quarks and gluons are badly %%@
constrained. Gluons are constrained around the region $x\sim 0.03-0.04$ where $R_G(x,Q_0^2)\sim 1$:
If gluon shadowing (see Fig. \ref{Fig:Initial}) at $x\sim 0.01...0.03$ were clearly stronger than %%@
that of sea quarks (which in turn is constrained by the DIS and DY data through the DGLAP evolution), %%@
then the $\log Q^2$-slopes caused by the DGLAP evolution at $Q_0^2$ would become negative %%@
\cite{Eskola:2002us}, and this would be in a clear contradiction with the NMC data for the $Q^2$ %%@
dependence in Fig.~\ref{Fig:RF2AC}. Thus, the three smallest-$x$ panels in Fig.~\ref{Fig:RF2AC} serve %%@
as the best constraint one currently obtains from DIS for nuclear gluons. At smaller $x$, where no %%@
high-$Q^2$ DIS data exist, again both sea quark and gluon modifications are badly constrained and %%@
remain specific to the parametric form chosen. Therefore, the uncertainty bands given in %%@
Fig.~\ref{Fig:Errors} are to be taken as lower limits for the true uncertainties.

Regarding the gluon shadowing in Fig. \ref{Fig:Initial}, we should also emphasize that like in %%@
\emph{EKS98} the sea quark and gluon shadowings become the same by construction rather than as a %%@
result of unbiased $\chi^2$ minimization: As the DIS data practically only constrains gluons at %%@
$x\sim 0.03-0.04$, momentum conservation alone is not able to fix the height or location of the %%@
antishadowing peak in $R_G^A$ in such a way that a clear enough minimum in $\chi^2$ would be %%@
obtained. Therefore, and also to test the \emph{EKS98} framework, we set the limits of $y_a$ and %%@
$p_{y_a}$ such that $R_G^A\approx R_S^A$ at $x\rightarrow 0$. We nevertheless observed that the  %%@
$\chi^2$ minimization tended to decrease the amount of gluon (anti)shadowing rather than support a %%@
stronger (anti)shadowing. We have also tested that if we keep the negligible gluon modifications at %%@
$x\sim 0.03-0.04$ but double the gluon shadowing at $x\ll 0.01$ (Fig.~\ref{Fig:Errors}, the green %%@
line)  the overall quality of the fits is not much deteriorated, $\chi^2/N=0.95$, even if the quark %%@
sector is not changed at all and no further $\chi^2$ minimization is made. This demonstrates that the %%@
indirect constraints given by the DIS and DY data and the momentum sum rules for $R_G^A$ are not very %%@
stringent, and that further constraints are certainly necessary for pinning down the nuclear gluon %%@
distributions.

Table \ref{Table:Khi2} summarizes the $\chi^2$ values obtained in the previous global analyses
for the nPDFs. A more detailed comparison is presented in \cite{EKPS}.
We conclude here that the old \emph{EKS98} analysis resulted in a fit whose quality 
is as good as in the automated analyses of the present work \cite{EKPS}, and also that the $\chi^2/N$ %%@
we obtain is close to that in \emph{nDS} and somewhat smaller than in \emph{HKM} and \emph{HKN}. %%@
Note, however, that the data sets included in each analysis are not identical. Interestingly, the NLO %%@
analysis of \emph{nDS} seems to give the best $\chi^2/N$ so far. 

Based on Fig.~\ref{Fig:Errors} and on the equally good overall quality of the fits obtained, we also %%@
conclude that the old \emph{EKS98} results agree quite nicely with our results from the automated %%@
$\chi^2$ minimization \cite{EKPS}: see the red lines for \emph{EKS98} in Fig.~\ref{Fig:Errors}. Thus, %%@
there is no need for releasing a new LO parametrization for the nPDFs, the \emph{EKS98} works still %%@
very well. To improve our analysis in the future, however, we plan to include RHIC d+Au data (see the %%@
discussion below) as further constraints and also eventually extend the analysis to NLO pQCD.

\begin{table}[tbh]
\begin{center}
\begin{tabular}{llcccccc}
 Set  &Ref.		&$Q_0^2/$GeV$^2$	& $N_{\mathrm{data}}$	& $N_{\mathrm{params}}$	&  $\chi^2$	& %%@
$\chi^2/N$& $\chi^2/$d.o.f.\\
\hline
\hline
 This work 	 &\cite{EKPS}				 &1.69 	& 514	& 16	& 410.15	& 0.798		& 0.824\\
 {\em EKS98} &\cite{Eskola:1998iy}		 &2.25	& 479	& --	& 387.39	& 0.809		& --  \\ 
 {\em HKM}   &\cite{Hirai:2001np}		 &1.0	& 309	& 9		& 546.6		& 1.769		& 1.822\\
 {\em HKN}   &\cite{Hirai:2004wq}		 &1.0	& 951	& 9     &1489.8		& 1.567		& 1.582\\
 {\em nDS}, LO  &\cite{deFlorian:2003qf} & 0.4  & 420	& 27	& 316.35	& 0.753		& 0.806\\
 {\em nDS}, NLO &\cite{deFlorian:2003qf} & 0.4  & 420	& 27	& 300.15	& 0.715		& 0.764\\
 \hline
\end{tabular}                                                  \\

\caption[]{\small The overall qualities of the fits obtained in different global analyses of nPDFs. 
}
\label{Table:Khi2}
\end{center}
%\vspace{-0.5cm}
\end{table}

\section{Stronger gluon shadowing?}

\begin{figure}[!]
\vspace{0.4cm}
\centering\includegraphics[width=12cm]{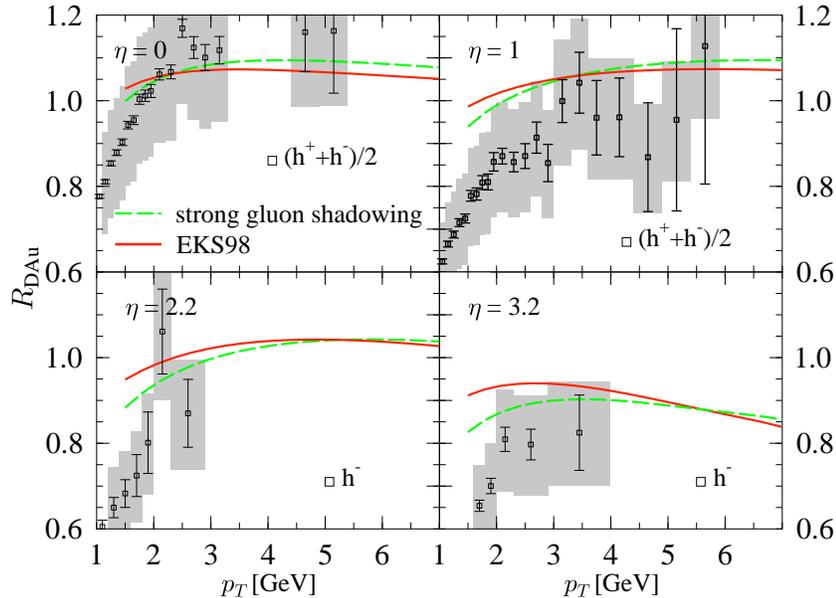}
\vspace{-0.2cm}
\caption[]{\small 
Minimum bias inclusive hadron production cross section in d+Au collisions divided 
by that in p+p collisions at $\sqrt{s}_{NN}=200$ GeV at RHIC as a function of hadronic transverse %%@
momentum at four different pseudorapidities.
The BRAHMS data \cite{Arsene:2004ux} are shown with statistical error bars and shaded systematic %%@
error bands.  A pQCD calculation for $h^++h^-$ production with the {\em EKS98} nuclear modifications %%@
and KKP fragmentation functions is shown by the red solid lines and that with the strong gluon %%@
shadowing of Fig.~\ref{Fig:Errors} by the dashed green lines.
}
\label{Fig:Brahms}
\vspace{-1.2cm}
\end{figure}

Further data sets to be included in the global analysis of nPDFs in the future, 
are provided by the d+Au experiments at RHIC. Figure \ref{Fig:Brahms} with BRAHMS data 
\cite{Arsene:2004ux} shows the ratio of inclusive $p_T$ distributions of hadrons 
at different pseudorapidities in d+Au collisions at $\sqrt s=200$ GeV  over those in p+p collisions.
The corresponding QCD-factorized LO cross sections are of the form
\begin{equation}
\sigma^{AB\rightarrow h+X} = \sum_{ijkl} 
f_i^A (x_1,Q) \otimes f_j^B(x_2,Q) \otimes \sigma^{ij\rightarrow kl} 
\otimes D_{k\rightarrow h+X}(z,Q_f), 
\label{Eq:Fragmentation}
\end{equation} 
with $h$ $(k)$ labeling the hadron (parton) type.  The fragmentation functions $D_{k\rightarrow %%@
h+X}(z,Q_f)$ we take from the KKP LO set \cite{Kniehl:2000fe}. We set the factorization scales $Q$ %%@
and $Q_f$ to the partonic  and hadronic transverse momentum, correspondingly, and define $z=E_h/E_k$ %%@
as the fractional energy. 

To test the sensitivity of the computed inclusive cross sections to gluon shadowing, we compute the 
cross sections by taking the nuclear modifications of PDFs from \emph{EKS98}
and from present analysis supplemented with the stronger gluon shadowing in Fig.~\ref{Fig:Errors}.
Note, however, that the systematic error bars in the BRAHMS data are large, and also that at the %%@
largest rapidities the data stand for negative hadrons only, while the KKP gives an average %%@
$h^++h^-$, and that we have not tried to correct for this difference in the computation. In any case, %%@
the large-$\eta$ BRAHMS data seems to suggest a stronger gluon shadowing than the relatively weakly %%@
constrained modest gluon shadowing obtained on the basis  of DIS and DY data in the global nPDF %%@
analyses. Too see whether such a strong gluon shadowing can be accommodated in the DGLAP framework
without deteriorating the good fits obtained, a careful global reanalysis must, however, 
be performed. In particular, it will be interesting to see whether changes in the gluon shadowing %%@
induce changes in the quark sector in such a way that the good agreement with the measured $\log Q^2$ %%@
slopes in Fig.~\ref{Fig:RF2AC} could be maintained.

\end{document}